\documentclass{emulateapj}

\shorttitle{Expected gamma-ray emission of SN~1987A}
\shortauthors{E.G. Berezhko, L.T. Ksenofontov \& H.J. V\"olk}

\newcommand{\gr}{$\gamma$-ray \,}

\begin{document}

\title {Expected gamma-ray emission of supernova remnant SN~1987A}

\author{E.G.~Berezhko\altaffilmark{1},
        L.T.~Ksenofontov\altaffilmark{1},
and     H.J.~V\"olk\altaffilmark{2}
}
\altaffiltext{1}{Yu.G. Shafer Institute of Cosmophysical Research and Aeronomy,
                 31 Lenin Ave., 677980 Yakutsk, Russia}
\altaffiltext{2}{Max Planck Institut f\"ur Kernphysik,
                Postfach 103980, D-69029 Heidelberg, Germany}

\email{ksenofon@ikfia.ysn.ru}

\begin{abstract}
A nonlinear kinetic theory of cosmic ray (CR) acceleration in supernova remnants
is employed to re-examine the nonthermal properties of the remnant of SN~1987A
for an extended evolutionary period of 5--100 yr. It is shown that an efficient
production of nuclear CRs leads to a strong modification of the outer supernova
remnant shock and to a large downstream magnetic field $B_\mathrm{d}\approx
20$~mG. The shock modification and the strong field are required to yield the
steep radio emission spectrum observed, as well as to considerable
synchrotron cooling of high energy electrons which diminishes their X-ray
synchrotron flux. These features are also consistent with the existing X-ray
observations. The expected \gr energy flux at TeV-energies at the current epoch
is nearly $\epsilon_{\gamma}F_{\gamma}\approx 4\times 10^{-13}$~erg
cm$^2$s$^{-1}$ under reasonable assumptions about the overall magnetic field
topology and the turbulent perturbations of this field. The general nonthermal
strength of the source is expected to increase roughly by a factor of two over
the next 15 to 20 yrs; thereafter it should decrease with time in a secular
form.
\end{abstract}



\section{Introduction}
The Supernova (SN) which occurred in 1987 in the nearby Large Magellanic Cloud
was the first object of its kind in modern times whose evolution could be
spatially resolved as a function of time; this includes the characterization of
the progenitor star. SN~1987A has been extensively studied in all wavelengths
from the radio to the \gr range. For a summary of its global characteristics
see  e.g. \citet{mccray93}.

The properties and the future evolution of this supernova can be viewed from
different physics points of view. The present work concentrates on the
nonthermal characteristics and, in particular, on the particle acceleration
aspects and the expected \gr emission. It extends two earlier studies by
\citet{bk00}, and by \citet{bk06}, hereafter referred to as BK, to include the
subsequent Chandra observations of the soft and hard X-ray emission
\citep{park07}, as well as the most recent and very detailed observations of the
radio continuum emission by \citet{zanardo10}. At the same time a qualitative
and in fact semi-quantitative prediction of the future nonthermal emission from
this source for the next decades is attempted.

The radio synchrotron emission and its spectral index are important for the
determination of the nonthermal energy density in the source.  In particular
the softness of the observed radio synchrotron spectrum shows that the radio
electrons are accelerated in a rather weak shock. This is the plasma subshock
formed when the accelerated nuclei modify the supernova remnant (SNR) blast
wave by their pressure gradient, so that the main gas compression occurs in an
extended CR precursor that does not accelerate radio electrons. This nonlinear
modification determines the injection rate of the nuclear particles and vize
versa (see BK and section 3).

The new Chandra data in addition suggest that the hard X-ray emission is
predominantly nonthermal. Since at the corresponding electron energies
synchrotron losses are of overriding importance for the shaping of the X-ray
synchrotron spectrum these new observations also determine the effective
magnetic field strength in the source \citep[see][for a review]{ber08}. 
It turns out that these recent observations can be consistently explained with
only minor modifications of the previous theoretical considerations. In
particular in can be shown that the circumstellar environment will lead to a
further increase of the expected \gr emission in the next decades by a modest
factor of only about 2, before the source will go into a steady emission
decrease as the SNR shock leaves the immediate circumstellar environment and
propagates secularly into the progenitor star's unperturbed Red Supergiant (RSG)
wind region.

In comparison with the other well-studied young, resolved SNR in the early
sweep-up phase, which also exhibits nonthermal X-ray emission and a steep radio
synchrotron spectrum, the Galactic object G1.9 + 0.3 that is presumably the
remnant of a type Ia SN in an essentially uniform interstellar medium (ISM),
SN~1987A has a complex immediate circumstellar environment due to the
interactions of its latest wind phases and the stellar radiation field before
the explosion. This leads to a very non-uniform density structure interior to
the stellar wind bubble formed by the progenitor in its RSG phase. Therefore
G1.9 + 0.3 and SN~1987A presumably constitute the extremes of spatially
resolved pre-supernova circumstellar environments when studying the particle
acceleration properties of extremely young SNRs. To this extent the present
work is complementary to the analysis of the nonthermal properties of G1.9 +
0.3 \citep{kvb10}. As one might expect on physics grounds, it turns out that in
both cases the SN blast wave is already strongly modified by the accelerated
particles -- mainly atomic nuclei, called here cosmic rays (CRs) --- from the
very beginning of SNR formation, although the total amount of nonthermal
energy, like that of thermal energy, is still only a percent fraction of the
total explosion energy. This shock modification is associated with a strong
amplification of the magnetic field strength in the swept-up circumstellar
material. These conclusions are derived from the following observations: (i)
the existence of nonthermal X-ray emission (see section 4), (ii) the rapid
increase of the radio emission \citep[see][for a recent analysis of the radio
light curve for G1.9 + 0.3]{murphy08}, and (iii) the anomalously soft radio
emission spectra (BK,~\citet{kvb10}). In both cases these evidences come from
the observed properties of synchrotron emission produced by the electron CR
component.  The electron component therefore plays a fundamental role,
permitting indirect conclusions about key properties of the nuclear energetic
particle component, like its pressure and its amplification of the magnetic
field. Given the diffuse radiation field, the inverse Compton gamma-ray
emission by the electrons is then also determined, and a measurement of the
total gamma-ray emission yields the hadronic gamma-ray component.

SN~1987A and G1.9+0.3 are nevertheless distinct from other young historic
Galactic SNRs like the core collapse object RX J1713.7 + 3946 and the type Ia
object SN~1006, respectively, in that the latter SNRs have already swept up an
amount of circumstellar matter that is approximately equal to the ejected mass.
Therefore they are expected by a number of authors to already contain an
amount of nonthermal particle energy that corresponds to some 10 \% of the
total hydrodynamic explosion energy
\citep{aha06,bkv09,bv10,acero10,zirak10,ell10}.

As a consequence of the softer
  effective equation of state of the combined gas-CR system the dynamical
  evolution in this phase is different from that of a purely gas dynamic
  system. Also the increasing dissipation of hydromagnetic fluctuation fields
  will presumably influence the system \citep{pz03,pz05}, not to speak of an
  early onset of radiative gas cooling as a result of the increased gas
  compression in the interior that is higher than the purely gas dynamic ratio
  of 4 \citep[][]{dorfi91}.

  The observed shape of the remnant, especially at radio frequencies is not
  very far from spherically symmetric, even though the observed emission varies
  around the remnant's periphery \citep{potter09}.  This suggests that a
  spherically symmetric model is a reasonable first order approximation for a
  study of the general nonthermal properties of the remnant.

  In the next section an approximate spherically symmetric model of the
  pre-explosion environment of SN~1987A is formulated.  This is combined in
  section 3 with a theoretical, spherically symmetric, nonlinear model for the
  nonthermal evolution, using the earlier work of BK, and with the recent
  observations of the synchrotron emission, mentioned before. The results are
  presented in section 4 and summarized in section 5.

\section{Approximate spherically symmetric model of the circumstellar
 environment SNR 1987A} 

The initial short outburst of radio emission \citep{turtle87} has been
attributed to the synchrotron emission of electrons accelerated by the SN shock
propagating in the free wind of the blue supergiant (BSG) progenitor star
\citep{chf87}. After about 3 years radio emission was detected again
\citep{ss92, gae97}, as well as a monotonically increasing X-ray emission
\citep{gro94, has96}. This second increase of emission has then been attributed
to the entrance of the outer SN shock into the wind bubble of the BSG of density
$\rho_\mathrm{B}$, thermalized in a termination shock at a radial distance
$R_\mathrm{T}=3.1\times 10^{17}$~cm \citep[e.g.][]{bk00}, and subsequently into
an H~II region of much denser matter, consisting of the swept-up wind of a RSG
precursor phase \citep{chd95}. The density $\rho_\mathrm{R}$ of the swept-up RSG
wind region is large compared to $\rho_\mathrm{B}$. The contact discontinuity
between the two winds is assumed at $R_\mathrm{C}=5\times 10^{17}$~cm, the scale
$l_\mathrm{C}$ of a modeled smooth transition is assumed as
$l_\mathrm{C}=0.05R_\mathrm{C}$.

These parameters are consistent with canonical values of stellar ejecta mass
$M_\mathrm{ej}=10M_{\odot}$, distance $d=50$~kpc, and hydrodynamic explosion
energy $E_\mathrm{sn}=1.5\times 10^{51}$~erg \citep[e.g.][]{mccray93}. During an
initial period the ejecta material has a broad distribution in velocity $v$. The
fastest part of this distribution can be described by a power law
$dM_\mathrm{ej}/dv\propto v^{2-k}$. The present modeling uses a value $k=8.6$
appropriate for SN~1987A \citep{mccray93}. The general picture is that the
interaction of the ejecta with the circumstellar medium (CSM) creates a strong
shock there which heats the thermal gas and accelerates particles diffusively to
a nonthermal CR component of comparable energy density.

According to \citet{chd95} the CSM at $r>R_\mathrm{C}$ includes three subsequent
regions: the H~II region, an equatorial ring and, beyond the ring, a free RSG
wind (followed further out by a rarefied main sequence wind bubble which is of
no concern here). Neither the contact discontinuity between the two winds is in
reality a spherical surface at radius $R_\mathrm{C}$, nor is the equatorial ring
a spherical shell, as the optical observations clearly indicate
\citep{mccray93,chd95}. Nevertheless, the present nonthermal model approximates
the mass distribution $\rho_\mathrm{ER}$ within the ring by a spherically
symmetric shell
\begin{equation}
\rho_\mathrm{ER}=\rho_\mathrm{m} \exp[-(r-R_\mathrm{R})^2/l_\mathrm{R}^2],
\end{equation}
where $\rho_\mathrm{m}\approx M_\mathrm{sh}/(4\pi^{3/2}
R_\mathrm{sh}^2l_\mathrm{R})$ is the central (maximal) density of the shell; $
M_\mathrm{sh}$, $R_\mathrm{R}$ and $l_\mathrm{R}$ represent the total mass, the
radius and the width of the shell, respectively. Below, the values
$R_\mathrm{R}=7\times10^{17}$~cm and $l_\mathrm{R}=0.25R_\mathrm{R}$ are used.
Approximately $l_\mathrm{R}$ corresponds to the interval of radial distances
where the dense cool shell, associated with the equatorial ring, is positioned
around the source \citep{crots95}.

In the same spirit, the density profile of the RSG matter consists of three
corresponding terms:
\begin{equation}
\rho_\mathrm{R}=\rho_\mathrm{II}H(R_\mathrm{R}-r) +
\rho_\mathrm{ER}+\rho_\mathrm{w} H(r-R_\mathrm{R}).
\end{equation}
Here $H(x)$ is the Heaviside function.
 
This results in the following model for the CSM density distribution at the
distances $r>R_\mathrm{T}$ in the form
\begin{equation}
\rho V_{0}=\frac{\rho_\mathrm{B}+\rho_\mathrm{R}}{2}-
\frac{\rho_\mathrm{B}-\rho_\mathrm{R}}{2}
\tanh\frac{r-R_\mathrm{C}}{l_\mathrm{C}},
\end{equation}

For the gas number density $N_\mathrm{g}=\rho/m_\mathrm{p}$ the following
values are adopted:
$N_\mathrm{g}^B=0.29$~cm$^{-3}$, $N_\mathrm{g}^\mathrm{II}=280$~cm$^{-3}$,
where $m_\mathrm{p}$ is the mass of proton. These values, as was shown by BK,
provide a good compromise between the SN shock dynamics seen in the radio and
X-ray emissions.

Since the density of the free RSG wind can be expressed in the form
\begin{equation}
\rho_\mathrm{w}= \frac{\dot{M}}{4\pi V_\mathrm{w} r^2},
\end{equation}
where $\dot{M}$ is the RSG mass loss rate and $V_\mathrm{w}$ denotes the RSG
wind speed,
the mass of the shell can be expressed in the form
\begin{equation}
M_\mathrm{sh}=\frac{\dot{M}R_\mathrm{R}} {V_\mathrm{w}} -\frac{4\pi
\rho_\mathrm{II}(R_\mathrm{R}^3-R_\mathrm{C}^3)}{3}.
\end{equation}
It is assumed that the center of the spherically smoothed ring is positioned
exactly at the boundary $r=R_\mathrm{R}$ between the HII region and the free RSG
wind region.

According to \citet{bl93} the RSG mass loss rate $\dot{M}$ is highly nonuniform:
it is $\dot{M}=8\times 10^{-5}M_{\odot}$~yr$^{-1}$ in the equatorial region
within $\pm 9^{\circ}$, and $\dot{M}=4\times 10^{-6}M_{\odot}$ yr$^{-1}$ in all
other directions. In the present spherically symmetric approximation its average
value is then $\dot{M}=1.6\times 10^{-5}M_{\odot}$ yr$^{-1}$.

For the conventional value $V_\mathrm{w}=5$~km s$^{-1}$ of the RSG wind speed
the masses of the
shell is $M_\mathrm{sh}=0.5\times M_{\odot}$.

\section {Particle acceleration model}
To describe the observed properties of nonthermal emission of SN~1987A a
nonlinear kinetic theory is used here. It couples the particle acceleration
process with the hydrodynamics of the thermal gas \citep{byk96, bv00}.
Therefore, in a spherically symmetric approach it is able to predict the
evolution of gas density, pressure, mass velocity, together with the energy
spectrum and the spatial distribution of CR nuclei and electrons at any given
evolutionary epoch $t$, and the properties of the nonthermal radiation produced
in SNRs due to these accelerated particles.

The model does not contain the reverse shock propagating in the SN ejecta.  The
reverse shock is not expected to be an efficient CR accelerator, because the
magnetic field value is expected to be very weak as a result of the large
expansion of the exploded stellar material that leads to inefficient high
energy CR acceleration.  Even if one assume equally efficient CR
injection/acceleration on the forward and the reverse shock, the expected
energy content of CRs accelerated by the forward shock is considerably larger
compared with the reverse shock except during the transition phase from the
free expansion to the Sedov SNR phase \citep[see][for more
details]{bk00}. Since SN~1987A is still far from the Sedov phase one can
neglect the contribution of the reverse shock to the nonthermal emission of the
remnant.  In addition the analysis of the observed fine structure of the X-ray
and the radio emission \citep{zhekov10} gives evidence that they are produced
by the forward shock (blast wave).

In the particle acceleration model the scattering properties of the CSM are
required. Given the enormous shock velocities in excess of $\approx 5000$~km
s$^{-1}$ (see next section), Bohm diffusion appears a most reasonable
approximation.
It was also used by BK. In this limiting case the scattering mean free path
equals the particle gyro radius and is therefore inversely proportional to the
mean magnetic field strength $B$.

A rather high downstream magnetic field strength $B_\mathrm{d}\approx 10$~mG is
required to reproduce the observed radio and X-ray spectra (BK). It is far from
obvious that the dynamical interaction of the RSG and BSG wind systems,
responsible for the inferred circumstellar structure, can also lead to such high
magnetic field strengths. It is much more likely that the required strength of
the magnetic field has to be attributed to nonlinear field amplification at the
SN shock by the CR acceleration process itself. According to plasma physical
considerations going back to \citet{lb00} and \citet{bell04}, the existing CSM
magnetic field can indeed be significantly amplified at a strong shock by CR
streaming instabilities. In fact, for all the thoroughly studied young SNRs,
the ratio of magnetic field energy density $B_0^2/8\pi$ in the upstream region
of the shock precursor to the CR pressure $P_\mathrm{c}$ is about the same
\citep{vbk05}. Here $B_0$ is the far upstream field in the precursor, presumably
amplified by the energetically dominant 
CRs of the highest energy. Within an error of about 50 percent the empirical
relation
\begin{equation}
B_0=\sqrt{2\pi \times 10^{-2}\rho_0 V_\mathrm{s}^2}
\end{equation}
holds, where $V_\mathrm{s}$ denotes the SN shock speed. This is quantitatively
close to the above-mentioned field value.  This is the same field value as used
by BK.  The factor $10^{-3}$ instead of $10^{-2}$ in Eq.(4) of BK is simply a
misprint: this is clearly seen from a comparison of Fig.1 of BK with Fig.1 of
the present paper.


For the following it is assumed that this relation also holds for an individual
young source like SN~1987A during its temporal evolution, where
$V_\mathrm{s}(t)$ depends on the source age $t$. For time scales of the order of
the propagation time of the shock through the shock precursor, i.e. about $10
\%$ of the system age \citep[e.g.][]{byk96} this appears as a reasonable
approximation.

As reviewed earlier \citep{voelk04, ber05, ber08}, and elaborated most recently
in detail \citep{bkv09,kvb10}, the key parameters of the theoretical model
(proton injection rate $\eta$ and electron to proton ratio $K_\mathrm{ep}$ below
the
synchrotron cooling range) can be determined in a semi-empirical way from a fit
of the nonlinear theoretical solution to the observed synchrotron emission
spectrum. For the sake of simplicity the values of these parameters are usually
assumed to be constant during SNR evolution. Also BK proceeded in this way. The
present paper assumes such a simplification to be the main reason why a
corresponding theoretical model fits the data only on average. This concerns the
shape of the radio spectrum and the time-dependence of the radio flux over an
extended period of SNR evolution \citep{zanardo10}. In the case of SN~1987A, on
the other hand, it is natural to expect substantial time-variations of these
injection parameters since the SN shock propagates through a strongly nonuniform
CSM whose physical parameters at the shock front are expected to change in time.
 Therefore, in contrast to BK, time dependent injection parameters $\eta(t)$ and
$K_\mathrm{ep}(t)$ will be admitted here. Their values are determined from the
fit to the measured synchrotron data. The main physical factor which determines
the injection efficiency $\eta(t)$ is the structure of the magnetic field
upstream of the shock. Injection is expected to be progressively less efficient
when the magnetic field component tangential to the shock surface becomes more
relevant \citep{vbk03}. Such a situation is expected in the most compressed CSM
region, that is, in the ring region. It is important to note that a reliable
semi-empirical estimate for the values of these parameters for any given SNR
evolutionary phase is possible if the measured synchrotron spectra in the radio
and X-ray bands are available for this particular phase
\citep[e.g.][]{bkv09,kvb10}. For SN~1987A good quality measurements of the radio
spectrum, as well as of the X-ray fluxes in the soft (0.5--2 keV) and the hard
(3--10 keV) energy ranges exist now for the full evolutionary period
\citep[e.g.][]{zanardo10}. The remaining problem is that the nature of the
observed X-ray emission of SN~1987A is not known unequivocally. In the extreme
case, when only an upper limit for the nonthermal (synchrotron) emission is
known, the data yield only a lower limit for the magnetic field strength
$B_0(t)$ and a corresponding value of the electron to proton ratio (BK). It is,
however, noted here that the time dependence of the hard X-ray flux differs from
that of the soft X-ray flux, and that it is very close to the time dependence of
the radio emission flux. This fact can be interpreted as evidence that the hard
X-ray emission is predominantly of a nonthermal nature
\citep[e.g.][]{zanardo10}. The considerations below are based on this
interpretation. 

The present model calculations start at the SNR evolutionary epoch $t=1000$~d,
when the outer SN shock has reached a radius $R_\mathrm{i}=R_\mathrm{T}$ and a
speed $V_\mathrm{i}=28000$~km s$^{-1}$. These values of $R_\mathrm{s}$ and
$V_\mathrm{s}$ correspond to the end of the SN shock propagation in the free BSG
wind region $r<R_\mathrm{T}$ (BK). The contribution of CRs accelerated in the
region $r<R_\mathrm{T}$ is neglected, because the number of CRs produced in the
region $r>R_\mathrm{T}$ becomes dominant very soon, because of the high gas
density there.
\begin{figure}
\plotone{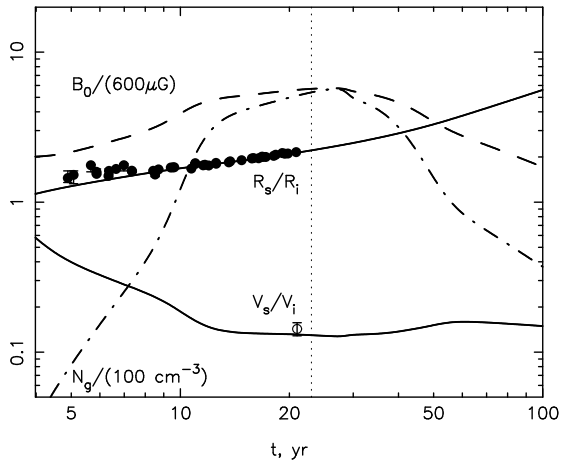} 
\figcaption{Shock radius $R_\mathrm{s}$ (solid), shock speed $V_\mathrm{s}$
(solid), gas density $N_\mathrm{g}$ (dash-dotted) and upstream magnetic field
$B_0$ (dashed) at the current shock position as a function of time since SN
explosion for $R_\mathrm{R}=7\times 10^{17}$~cm. The {\it dotted vertical line}
marks the current epoch. The observed radius and speed of the SN shock, as
determined by radio observations \citep{ng08} are shown as well. The scaling
values are $R_\mathrm{i}=R_\mathrm{T}=3.1\times 10^{17}$~cm and
$V_\mathrm{i}=28000$~km~s$^{-1}$
\label{f1}}
\end{figure}

\section{Results and Discussion}

The calculated shock radius $R_\mathrm{s}$ and shock speed $V_\mathrm{s}$, shown
in Fig.\ref{f1} as a function of time, are in satisfactory agreement with the
values obtained on the basis of radio measurements. Up to the year 2030 the
shock speed decreases due to the increase of the CSM density and then, after
crossing the maximal CSM density, it increases again from $V_\mathrm{s}=3300$~km
s$^{-1}$ to $V_\mathrm{s}\approx 4400$~km s$^{-1}$. The reason for this
unexpected behavior is that during the whole time period under consideration
the swept-up mass is much lower than the ejecta mass $M_\mathrm{ej}$. Therefore
the main fraction of the explosion energy is contained in the freely expanding
ejecta. On a small time scale the shock is driven by the downstream
overpressure: each local relatively sharp increase of the CSM density leads to a
temporal decrease of the shock speed that keeps $\rho_0V_\mathrm{s}^2$ constant.
However, on a larger time scale the shock is piston driven.

Note that the actual CSM is essentially non-spherically symmetric during the
time period considered. Despite of this situation the piston-driven shock is
expected to be roughly spherical. This supports the spherically symmetric
approach used here.

\begin{figure}[t]
\plotone{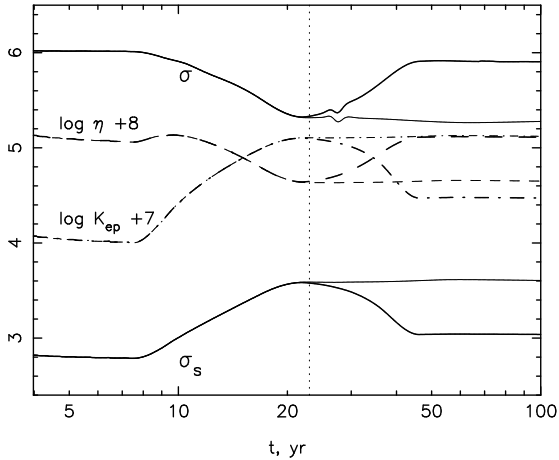}
\figcaption{Shock compression ratio $\sigma$ (solid), subshock compression
  ratio $\sigma_\mathrm{s}$ (solid), proton injection rate $\eta$ (dashed), and
  electron-to-proton ratio $K_\mathrm{ep}$ (dash-dotted) as a function of
  time. {\it The dotted vertical line} again marks the current epoch.
  The thick and the thin lines here and in the subsequent figures correspond to
  an increasing (= high) and to a constant (= low) proton injection rate after
  an age of $\sim 23$~yrs, respectively.
\label{f2}}
\end{figure}

The fit of the theoretical solution to the synchrotron spectra, as measured in
the radio and X-ray ranges up to the year 2008 \citep{zanardo10}, yields
estimates for the proton injection rate $\eta(t)$ and for the electron-to-proton
ratio $K_\mathrm{ep}(t)$. This is shown in Fig.\ref{f2}. The fit procedure was
described in detail for similar cases \citep{bkv09,kvb10}. The required proton
injection rate $\eta(t)\approx 10^{-3}$ leads to a significant nonlinear
modification of the shock: as can be seen in Fig.\ref{f2} the total shock
compression ratio $\sigma = 5.4-6$ is essentially larger, and the subshock
compression ratio $\sigma_\mathrm{s}\approx 2.8-3.6$ is lower than the classical
value of 4 for a pure gas shock. Fig.\ref{f2} also shows that the proton
injection rate $\eta$ changes during the SNR evolution, so that it has a local
minimum at an age of about $20$~yr. This is required in order to fit the
observed radio emission spectra, which are becoming somewhat harder than a
spectrum calculated with an unchanged injection rate - at least up to the
present time. During the period from $t \approx 13$~yr to $t \approx 30$~yr,
i.e. beyond the present epoch of radio observations, the shock is within the
dense shell that corresponds to the observed equatorial ring. In this spatial
region the magnetic field is expected to have an enhanced tangential component
which should depress the nuclear injection rate. 

Thereafter the injection rate is a priori unknown. It may increase again to the
same level $\eta(t)\approx 10^{-3}$.  This is one of the possibilities
calculated below (so-called high proton injection rate). However, since the
magnetic field vector in the free RSG wind is dominated by a component
tangential to the SNR shock also the possibility of a continuing depressed
injection (so-called low proton injection rate) at a level similar to the value
$\eta(t)\approx 4\times 10^{-4}$ within the ring region is considered.

\begin{figure}[t]
\plotone{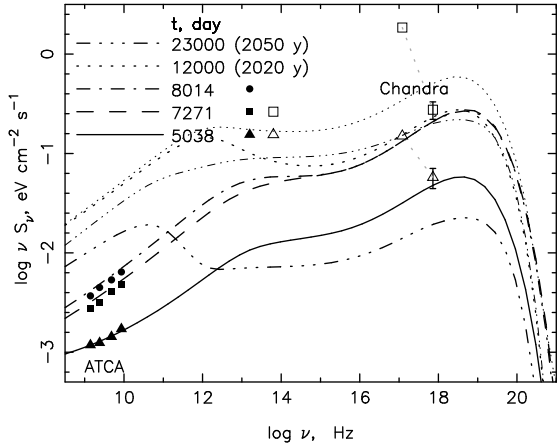} 
\figcaption{Synchrotron spectral energy density of SN~1987A, calculated for the
five evolutionary epochs ({\it solid curves}). The {\it ATCA} radio
\citep{zanardo10} and data for three epochs are shown as well, together with
Chandra X-ray \citep{park07} data for two epochs (both, in the soft energy
range at $\nu = 10^{17}$~Hz and, connected by {\it dotted lines}, in the
hard energy range at $\nu = 6 \times 10^{17}$~Hz).
\label{f3}}
\end{figure}

\begin{figure}[t]
\plotone{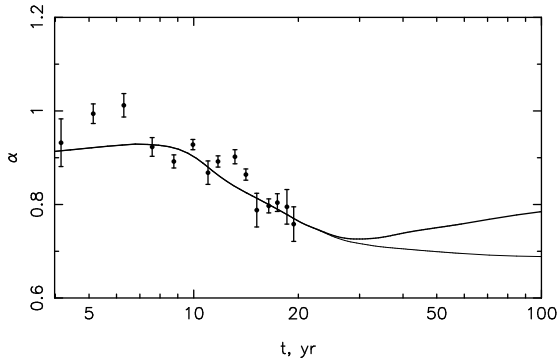} 
\figcaption{Spectral index of the radio synchrotron emission of SN~1987A as a
  function of time, together with the {\it ATCA} data
\citep{zanardo10}. 
\label{f4}} 
\end{figure}

The higher injection rate yields for $t>23$~yr a correspondingly larger shock
modification, characterized by a larger shock compression ratio $\sigma$ and
a lower subshock compression ratio $\sigma_\mathrm{s}$ (Fig.\ref{f2}).

The strongly modified SNR shock generates a CR spectrum $N\propto p^{-\gamma}$,
which is very soft at momenta $p<m_\mathrm{p}c$, with index $\gamma \approx
(\sigma_\mathrm{s}+2)/(\sigma_\mathrm{s}-1)\approx 2.7$. CR electrons with such
a spectrum produce a radio synchrotron emission spectrum $S_{\nu}\propto
\nu^{-\alpha}$ with spectral index $\alpha =(\gamma-1)/2\approx 0.9$, that
corresponds very well to the observations, as can be seen in Fig.\ref{f3}, where
the synchrotron spectral energy density $\nu S_{\nu}$ is calculated for five
successive epochs together with the experimental data.

The assumed time variation of the proton injection rate $\eta$ leads to a time
dependence of the radio spectral index $\alpha$ as shown in Fig.\ref{f4}, and
it provides a better fit of the observations \citep{zanardo10} compared with
the case of constant $\eta$ (BK).  In fact the adopted injection rate $\eta(t)$
at $t<20$~yr is the result of the fit to the radio 
synchrotron spectrum.

For the future, $t>23$~yr, the slope of the radio synchrotron spectrum is
  expected to be nearly the same if the proton injection remains constant at
  $\eta \approx 4\times 10^{-4}$ (Fig.\ref{f4}). It is expected to become
  progressively steeper due to the decrease of the subshock compression ratio
  $\sigma_\mathrm{s}(t)$ in case the proton injection rate increases
  again to the level $\eta \approx \times 10^{-3}$.

The strong downstream magnetic field $B_\mathrm{d}\approx 20$~mG leads to
synchrotron
cooling of the electrons with momenta $p>10m_\mathrm{p}c$. This makes the high
energy part of the synchrotron spectrum ( $\nu> 10^{13}$~Hz) very soft (see
Fig.\ref{f3}). At higher frequencies ( $\nu= 10^{16}-10^{19}$~Hz) the
synchrotron spectrum becomes harder, possibly due to a pile-up effect. It
hardens the spectrum of accelerated electrons, that undergo strong synchrotron
losses, just near its exponential cutoff \citep[e.g.][]{drury99}. Under this
condition the calculated synchrotron flux at frequency $\nu\approx 10^{17}$~Hz,
which corresponds to a photon energy $\epsilon_{\gamma}=0.5$~keV, is below the
measured flux (see Fig.\ref{f3}). This is a required condition because at
energies $\epsilon_{\gamma}=0.5-2$~keV the X-ray emission of SN~1987A is
dominated by lines and is therefore mainly of thermal origin. At higher
energies $\epsilon_{\gamma}> 3$~keV X-rays are presumably of {a predominant}
nonthermal origin. Therefore the fit of the measured X-ray flux for
$\epsilon_{\gamma}= 3$~keV was used in the determination of the values of the
injection parameters ($\eta(t)$ and $K_\mathrm{ep}(t)$) and of the amplified
magnetic field $B_0(t)$ at the beginning of the shock precursor.

The less modified shock corresponding to a low injection rate during the future
epoch $t>23$~yr yields lower cooling of electrons due to
the lower downstream magnetic field on account of the lower overall
compression ratio. This leads to a considerably flatter high-energy part of
the synchrotron spectrum ( $\nu> 10^{13}$~Hz).

\begin{figure}
\plotone{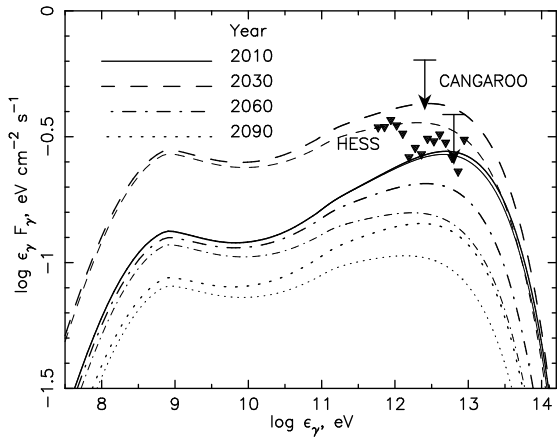} 
\figcaption{Integral \gr energy flux from SN~1987A, calculated for four
epochs. The CANGAROO ({\it arrows}) \citep{eno07} and the latest H.E.S.S.
({\it triangles}) \citep{komin10} upper limits are shown as well. 
\label{f5}}
\end{figure}

The calculated \gr integral spectral energy flux density (SED), shown in 
Fig.\ref{f5}, is dominated by the $\pi^0$-decay component at all energies.

It is important to note here that the hadronic SED has nevertheless been
renormalized by a factor $f_\mathrm{re}=0.2$ compared to the amplitude resulting
from the spherically symmetric model on which the entire calculation is
predicated \citep[e.g.][]{vbk03}. The reason for this a posteriori correction is
the fact that the circumstellar environment is characterized by a Archimedean
 spiral topology, both in the BSG as well as in the RSG wind bubble environment.
In particular the radiative cooling and the ionization by the stellar radiation
field of the progenitor star will make the entire interaction region of these
successive wind phases highly turbulent. Therefore there will always be regions
where the SNR is quasi-parallel, that is to say, regions where the shock normal
will make an relatively small angle with the local upstream magnetic field
direction. Only in these quasi-parallel shock regions nuclear particles can be
effectively injected into the diffusive acceleration process. However, as a
consequence of the overall Archimedean spiral topology, the total solid angle of
these quasi-parallel shock segments will be $4\pi f_\mathrm{re}$, where
$f_\mathrm{re}$ is considerably smaller than unity. Following previous estimates
\citep{vbk03,bpv03} $f_\mathrm{re}=0.2$ is used in this paper, but the actual
value of $f_\mathrm{re}$ may be even smaller. The corresponding uncertainty in
the amplitude of the \gr energy flux remains.

Since the SN shock is strongly modified the \gr spectrum at energies
$\epsilon_{\gamma}>0.1$~TeV is very hard: $F_{\gamma}\propto
\epsilon_{\gamma}^{-0.8}$. At the current epoch the expected \gr energy flux at
TeV-energies is about $\epsilon_{\gamma}F_{\gamma}\approx 4\times 10^{-13}$~erg
cm$^{-2}$ s$^{-1}$. and during the next 20 years it is expected to grow by a
factor of about two with a subsequent temporal decrease $F_{\gamma}\propto
R_\mathrm{s}^{-1}$ due to the rapid spatial decrease of the CSM density
$\rho_\mathrm{w} (R_\mathrm{s}) \propto R_\mathrm{s}^{-2}$ \citep[e.g.][]{bv00}.

At present there exist only upper limits for the TeV emission. They have been
obtained by the CANGAROO \citep{eno07} and H.E.S.S. \citep{komin10} instruments
(see Fig.\ref{f5}). The latest H.E.S.S. upper limit is quite close to the SED
estimated above. Even with the reservations formulated above, this is remarkable
and justifies in the view of the present authors a deep observation of this
object.

The expected time profile $F_{\gamma}(t)$ is sensitive to the radial profile of
the actual CSM density distribution. If the dense shell, which in the
spherically symmetric model represents the matter contained in the equatorial
ring, is situated at a larger distance, say at $R_\mathrm{R}=9\times10^{17}$,
the
peak of \gr flux is expected to occur 10 years later, with an amplitude that is
by a factor of 1.3 lower than in the former case.

It is clear from the above consideration that the proton injection rate $\eta$
(whose value influences the efficiency of CR production) can be estimated from
the observed shape of the radio emission. Since it is not possible to predict
the values of $\eta$ for the future evolutionary epochs, the presented
prediction for the corresponding gamma-ray emission is uncertain. Note however
that this kind of uncertainty is not very big due to the following
reasons. First of all, it is hard to expect that the shape of the radio
spectrum will be much more rapidly changing during the next 20-30 years than
during the previous epochs.  Therefore the predictions based on the recently
determined value of $\eta$ should be roughly valid also for this period of
time.  Secondly, the shape of the CR spectrum is strongly sensitive to the
injection rate at low energies $\epsilon < m_\mathrm{p}c^2$.  Since the radio
synchrotron emission is produced by electrons with such energies it is
sensitive to the expected injection rate (Fig.\ref{f3},\ref{f4}).  The
high-energy part of the CR spectrum $(\epsilon \gg m_\mathrm{p}c^2)$ is less
sensitive
to $\eta$: variations of $\eta$ by a factor of three at the
epoch $t>40$~yr (see Fig.\ref{f2}) only lead to a variation of the CR pressure
$P_\mathrm{c}/(\rho_0V_\mathrm{s}^2)\approx 1-\sigma_\mathrm{s}/\sigma$ by a
factor of 1.5. Since the CR
pressure is dominated by the highest energy CRs one should expect a
corresponding variation of the high energy gamma-ray emission. This is indeed
seen from Fig.\ref{f5}: for $t>40$~yrs the expected gamma-ray flux at energies
$\epsilon_{\gamma}=1-10$~TeV for low injection rate is lower by a factor $\sim
1.5$ compared with the case of high injection.

Therefore, taking into account the uncertainties of all relevant parameter
values we expect that the prediction for the TeV-emission is uncertain at best
by a the factor of two (Fig.\ref{f5}).

\section{Summary}
A kinetic nonlinear model for CR acceleration in SNRs has been applied in detail
to SN~1987A, in order to compare its results with observed properties. It is
found that quite reasonable consistency with most of the observational data can
be achieved.

The evidence for efficient CR production, leading to a strong modification of
the shock, comes from the radio synchrotron data. A proton injection rate of
size $\eta \approx 10^{-3}$ is required to produce a significant shock
modification that leads to the steep spectrum of energetic
electrons which fits the observed synchrotron spectrum very well. The condition
is an extremely high downstream magnetic field strength $B_\mathrm{d}\sim
10$~mG. Such a high field implies significant synchrotron losses of CR electrons
emitting nonthermal X-rays. This makes the high frequency part of the
synchrotron spectrum much softer and is consistent with the 
high-energy part of the X-ray spectrum, which is presumably of nonthermal
origin. Therefore the fit of the synchrotron spectrum gives a good estimate for
the CR injection parameters and makes it possible to calculate the expected \gr
flux in a spherically symmetric model. However, the basic tendency of the
magnetic field vectors to form an Archimedean spiral configuration in the
circumstellar wind interaction region requires a renormalization of the overall
\gr flux. This a posteriori reduction by a factor $f_\mathrm{re} = 0.2$ is
chosen following the previous analyses of other SNRs. This introduces an
acknowledged uncertainty in the predicted overall \gr flux.

Since the SN shock interacts with the cool shell which is the densest part of
the CSM, the $\pi^0$-decay $\gamma$-ray spectral energy flux density at the
current epoch is already quite high $\epsilon_{\gamma}F_{\gamma}\approx 4\times
10^{-13}$ ~erg cm$^{-2}$ s$^{-1}$ at energies $\epsilon_{\gamma}=0.1-10$~TeV.
Depending upon the details of the CSM distribution during the next 15--20 yr the
\gr flux is expected to increase roughly by a factor of two. 

The further temporal evolution of the gamma-ray emission corresponds to a
secular decrease because the gas density in the unperturbed Red Supergiant
Wind region decreases with radius $\propto r^{-2}$. Since the particle
injection into the shock acceleration process can not be well determined
for this later phase, the precise form of the emission decrease is not
well known.

The detection of $\gamma$-ray emission from SN~1987A would be a very important
element in a consistent picture for this SNR. In particular, it would give
evidence for efficient CR production followed by strong magnetic field
amplification for a core collapse supernova at a very early stage of evolution
of its remnant.

\acknowledgements 
This work has been supported in part by the Russian Foundation for Basic
Research (grants 09-02-12028, 10-02-00154), the Council of the President of the
Russian Federation for Support of Leading Scientific Schools (project no. 
NSh-3526.2010.2), the Ministry of Education and Science (contract
02.740.11.0248). The authors thank Dr. V.N. Zirakashvili for pointing out a
numerical inaccuracy in the calculation of the X-ray synchrotron spectra in an
early version of this paper. EGB acknowledges the hospitality of the
Max-Planck-Institut f\"ur Kernphysik, where part of this work was carried out.


\begin{thebibliography}{}

\bibitem[Acero et al.(2010)]{acero10}
Acero, F., et al. 2010, \aap, 516, A62

\bibitem[Aharonian et al.(2006)]{aha06}
Aharonian, F., et al. 2006, \aap, 449, 223

\bibitem[Bell(2004)]{bell04}
Bell, A. R. 2004, \mnras, 353, 550

\bibitem[Berezhko et al.(1996)]{byk96}
Berezhko, E. G., Elshin, V. K., \& Ksenofontov, L. T. 1996, JETP, 82, 1

\bibitem[Berezhko \& V\"olk(2000)]{bv00}
Berezhko, E.G. \& V\"olk, H.J. 2000, A\&A, 357, 183

\bibitem[Berezhko \& Ksenofontov(2000)]{bk00}
Berezhko, E. G., \& Ksenofontov, L. T. 2000, Astronomy Letters, 26, 639

\bibitem[Berezhko et al.(2003)]{bpv03} 
Berezhko, E. G., P\"uhlhofer, G., V\"olk, H. J. 2003, \aap, 400, 971

\bibitem[Berezhko \& Ksenofontov(2006)]{bk06}
Berezhko, E. G., \& Ksenofontov, L. T. 2006, \apjl,650, L59 (BK).

\bibitem[Berezhko(2005)]{ber05}
Berezhko, E.G. 2005, Adv. Space Res., 35, 1031

\bibitem[Berezhko(2008)]{ber08}
Berezhko, E.G. 2008, Adv. Space Res., 41, 429

\bibitem[Berezhko et al.(2009)]{bkv09} 
Berezhko, E.G., Ksenofontov, L.T., \& V\"olk, H.J., 2009, A\&A , 505, 169

\bibitem[Berezhko \& V\"olk(2010)]{bv10}
Berezhko, E.G. \& V\"olk, H.J. 2010, \aap, 511, A34

\bibitem[Blondin \& Lundqvist(1983)]{bl93}
Blondin, J.M. \& Lundqvist, P. 1993, \apj, 405, 337

\bibitem[Chevalier \& Dwarkadas(1995)]{chd95}
Chevalier, R. A. \& Dwarkadas, V. V. 1995, \apjl, 452, L45

\bibitem[Chevalier \& Fransson(1987)]{chf87}
Chevalier, R. A., \& Fransson, C. 1987, Nature, 328, 44

\bibitem[Crots et al.(1995)]{crots95}
Crots, A.P.S., Kunkel, W.E. \& Heathcote, S.R. 1995, \apj, 438, 724

\bibitem[Dorfi(1991)]{dorfi91}
Dorfi, E. A. 1991, \aap, 251, 597

\bibitem[Drury et al.(1999)]{drury99}
Drury, L.O., et al. 1999, \aap, 347, 370

\bibitem[Ellison et al.(2010)]{ell10}
Ellison, D.C., et al. 2010, \apj, 712, 287

\bibitem[Enomoto et al.(2007)]{eno07}
Enomoto, R., et al. 2007, \apj, 671, 1939

\bibitem[Gaensler et al.(1997)]{gae97}
Gaensler, B. M., Manchester, R. N., Staveley-Smith, L., et al. 
1997, \apj, 479, 845

\bibitem[Gorenstein et al.(1994)]{gro94}
Gorenstein, P., Hughes, J. P., Tucker, W. H. 1994, \apjl, 420, L25
	
\bibitem[Hasinger et al.(1996)]{has96}
Hasinger, G., Ashenbach, B. \& Tr\"umper, J. 1996, \aap, 312, L9

\bibitem[Komin et al.(2010)]{komin10}
Komin, N., et al. (H.E.S.S. Collaboration) 2010, at COSPAR Scientific
Assembly, Bremen, E19-0096-10 (Poster, Nr. Thu-299)

\bibitem[Ksenofontov et al.(2010)]{kvb10}
Ksenofontov, L. T., V\"olk, H. J., \& Berezhko, E. G. 2010, \apj, 714, 1187

\bibitem[Lucek \& Bell(2000)]{lb00}
Lucek, S. G. \& Bell, A. R. 2000, \mnras, 314, 65

\bibitem[McCray(1993)]{mccray93}
McCray, R. 1993, \araa, 31 175

\bibitem[Murphy et al.(2008)]{murphy08}
Murphy, T., Gaensler, B. M.\& Chatterjee, S.2008, \mnras,389, L23

\bibitem[Ng et al.(2008)]{ng08}
Ng, C.-Y., Gaensler, B. M., Staveley-Smith, L. et al. 2008,
\apj, 684, 481

\bibitem[Park et al.(2007)]{park07}
Park, S., Burrows, D. N., Garmire, G.P. et al. 2007,
in AIP Conf. Proc. 937, Supernova 1987A: 20 Years After
ed. S. Immler, K.W. Weiler, \& R. McCray (Melville, NY: AIP), 43

\bibitem[Potter et al.(2009)]{potter09}
Potter, T. M., Staveley-Smith, L., Ng, C.-Y. et al. 2009, \apj, 705, 261

\bibitem[Ptuskin \& Zirakashvili(2003)]{pz03}
Ptuskin, V. S. \& Zirakashvili, V. N. 2003, \aap, 403, 1

\bibitem[Ptuskin \& Zirakashvili(2005)]{pz05}
Ptuskin, V. S. \& Zirakashvili, V. N. 2005, \aap, 429, 755

\bibitem[Staveley-Smith et al.(1992)]{ss92}
Staveley-Smith, L., Manchester, R. N., Kesteven, M. J. et al. 1992,
Nature, 355, 147

\bibitem[Turtle et al.(1987)]{turtle87}
Turtle, A. J., Campbell-Wilson, D., Bunton, J. D., et al. 1987, Nature, 327, 38

\bibitem[V\"olk(2004)]{voelk04}
V\"olk, H.J. 2004, in {\it Frontiers of Cosmic Ray Science}, Proc. 28th
ICRC, Tsukuba, 8, 29 ff, ed. T, Kajita, Y. Asaoka, A. Kawachi, Y. Matsubara, \&
M. Sasaki (Tokyo, Japan, Universal Academy Press Inc.)

\bibitem[V\"olk et al.(2003)]{vbk03}
V\"olk, H. J., Berezhko, E. G., \& Ksenofontov, L. T. 2003, \aap 409, 563

\bibitem[V\"olk et al.(2005)]{vbk05}
V\"olk, H. J., Berezhko, E. G., \& Ksenofontov, L. T. 2005, \aap, 433, 229

\bibitem[Zanardo et al.(2010)]{zanardo10}
Zanardo, G. et al. 2010, \apj, 710, 1515

\bibitem[Zhekov et al.(2010)]{zhekov10}
Zhekov, S. A., Park, S., McCray, R., et al. 2010, \mnras,407, 1157

\bibitem[Zirakashvili \& Aharonian(2010)]{zirak10}
Zirakashvili, V. N. \& Aharonian, F.A. 2010, \apj, 708, 965
\end{thebibliography}
\end{document}